# Online SOC Estimation of Lithium-ion Battery Based on Improved Adaptive H Infinity Extended Kalman Filter

Jierui Wang[1], Wentao Yu[2], Guoyang Cheng[1], Lin Chen[1]

1. College of Electrical Engineering and Automation, Fuzhou University, Fujian 350108, China;

2. Fujian Nebula Electronics Co., Ltd, Fujian 350015, China

**Abstract:** For the battery management system of electric vehicle, accurate estimation of the State of Charge of Lithium-ion battery can effectively avoid structural damage caused by overcharge or over discharge inside the battery. Considering that the lithium-ion battery is a time-varying nonlinear system, which needs real-time State of Charge estimation, a joint algorithm of forgetting factor recursive least squares and improved adaptive H Infinity Extended Kalman Filter is proposed for online estimation of model parameters and state of charge. Firstly, Thevenin equivalent circuit model is built in Simulink of MATLAB R2021b, and the model parameters are estimated by forgetting factor recursive least square in real time. Secondly, the improved adaptive H Infinity Extended Kalman Filter is used to estimate State of Charge in true time. Finally, the feasibility of the algorithm is verified by two different lithium-ion battery conditions. The experimental results show that improved adaptive H Infinity Extended Kalman Filter has the highest and most stable State of Charge estimation accuracy than the other three comparison methods. The Root Mean Square Error and Mean Absolute Error are 0.6008 % and 0.3578 % under the Dynamic Stress Test condition, and 1.0068 % and 0.8721 % under the Federal Urban Driving Schedule condition respectively.

## 1 Introduction

With the rapid development of new energy vehicles, countries have increased their R&D investment and use of new energy vehicles. According to the global electric vehicle 2022 report [1], about 16.5 million electric vehicles have been put into use worldwide so far after decades of development. For new energy vehicles, the battery management system (BMS) is like the brain that needs to pay attention to the operation of the on-board battery at all times. Among them, the State of Charge (SOC) of lithium-ion battery is one of the key objects. For SOC, BMS needs to be estimated quickly, accurately and stably [2].

Due to the unmeasurable characteristics of SOC, it is necessary to estimate the SOC of lithium-ion battery by algorithm based on the measured operating data such as temperature, current and voltage. The commonly used methods are coulomb counting method, data fusion method and nonlinear state estimation method based on Kalman filter and its variants. In the case of inaccurate initial SOC, coulomb counting method [3] will cause error accumulation, and then gradually deviate from the exact value. Data fusion method [2] requires a large number of test data to train the estimation model, but this method relies heavily on the amount of data and data accuracy, and the training cycle is long. The nonlinear state estimation method [4] is gradually applied to BMS as the mainstream method of SOC estimation because of its simple structure, convenient design and strong robustness.

At present, domestic and foreign scholars have done a lot of in-depth research work on SOC estimation based on nonlinear state estimation method. Ref.[5] used Extended Kalman Filter (EKF) algorithm to estimate the battery model parameters and states estimation. The biggest feature of the algorithm is to estimate the battery SOC value through the first-order linearized nonlinear system. Based on the principle of robustness, Ref.[6] proposed an H Infinity EKF (HIEKF) method to estimate SOC. Compared with EKF, this method improves the SOC estimation accuracy about 0.19 %. In Ref.[7], an improved adaptive

EKF (IAEKF) algorithm is proposed. The maximum likelihood estimation function is introduced to update the length of the innovation error sequence in real time. Compared with AEKF, the SOC estimation accuracy is improved, but the introduced error sequence length threshold needs to be repeatedly tested to finally determine the appropriate value. In addition, other KF algorithms are also used for SOC estimation. Unscented Kalman Filter (UKF) avoids the disadvantage of low first-order linearization accuracy of EKF through unscented transformation, but it has higher parameter requirements. Cubature Kalman Filter (CKF) is based on the third-order spherical radial volume criterion. The SOC value is estimated by introducing a volume point, which requires higher parameter requirements than UKF [8].

Considering the stability and complexity of the algorithm, this paper analyzes the principle of HIEKF and AHIEKF on the basis of EKF, and proposes an improved AHIEKF (IAHIEKF) according to the defect of the updating criterion of noise covariance matrix of AHIEKF algorithm. Finally, the superiority of the proposed algorithm is analyzed by experiments.

## 2 Model Introduction and Parameters Identification

Establishing an accurate equivalent circuit model (ECM) is an important step to accurately estimate the SOC of lithium-ion battery. The current common ECMs include Rint model, Thevenin model, PNGV model, n-RC model and fractional order model [9]. At present, Thevenin model is widely used in engineering. Based on the analysis of Ref. [9], this paper will use Thevenin model to characterize the characteristics of lithium-ion battery.

### 2.1 OCV-SOC Curve Processing

Before model parameters identification, it is necessary to know the value of Open circuit voltage (OCV) in advance. Most of the processing methods are to fit the functional relationship between SOC and OCV through specific conditions, such as low-rate charge or discharge fitting method [10] and static charge or discharge fitting method [11]. The low-rate charge or discharge method is to charge or discharge slowly at a small rate current (such as 0.01C or 0.05C). The terminal voltage measured during the charge or discharge process is approximated as the OCV of the battery, and then the SOC and OCV are fitted. The standing charge or discharge method is each interval $n$% SOC (such as 5% or 10%) to stand for 1-2 hours. The voltage measured after standing is used as the OCV corresponding to the current SOC, then the values of OCV and SOC measured are fitted.

Although the above two methods can accurately obtain the functional relationship between OCV and SOC, they are undoubtedly time-consuming and laborious. Therefore, the OCV is identified as an unknown parameter, this approach does not need to design a specific operating condition to obtain the OCV-SOC curve. In Ref.[12], OCV and Thevenin model parameters are identified together, but this method adds an additional identification parameter, which will undoubtedly have a certain impact on the final model parameters identification results. Therefore, the process of this paper: Firstly, OCV is identified by forgetting factor recursive least square (FFRLS) method based on Rint model, secondly, OCV is substituted into Thevenin model as a known value to identify other parameters by FFRLS [13].

Fitting OCV and SOC Function Relationship with 6-order polynomial:

$$f(z)=k_6*z^6+k_5*z^5+k_4*z^4+k_3*z^3+k_2*z^2+k_1*z+k_0 \quad (1)$$

Where $z$ represents SOC. After identifying the value of OCV, the value of SOC can be obtained by the discrete ampere-hour integral method of Eq.2, and then the FFRLS algorithm is used to identify the function coefficients in Eq.1. where coefficient vector and data vector are:

$\Theta=[k_6\ k_5\ k_4\ k_3\ k_2\ k_1\ k_0]$, $\Phi=[z^6\ z^5\ z^4\ z^3\ z^2\ z\ 1]^T$.

$$SOC_{k+1}=SOC_k-(\eta*\Delta t*I_k)/Q_n \quad (2)$$

Among them, $\eta$ is the coulombic efficiency, $\Delta t$ is the sampling time, $I_k$ is the measured current at $k$-th, and $Q_n$ is the actual available capacity of the battery.

### 2.2 Thevenin Model Parameters Identification

In the Thevenin model (Fig.1), $R_0$ represents the battery internal resistance, $R_p$ represents the polarization internal resistance, $C_p$ represents the polarization capacitance, $U_{OCV}$ represents the open circuit voltage, $U_t$ represents the terminal voltage, $U_0$ represents the internal resistance voltage, $U_p$ represents the polarization voltage, $I$ represents the current, and define the discharge direction is negative.

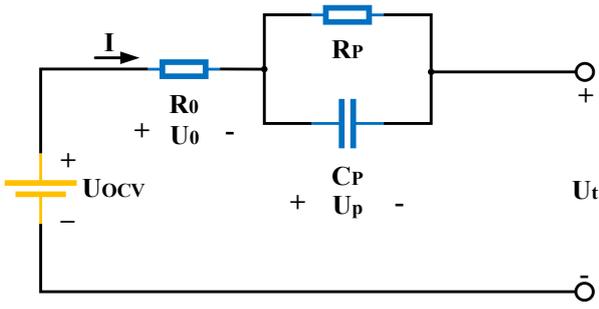

Fig. 1: The Thevenin equivalent circuit model

According to Kirchhoff voltage law:

$$dU_p/dt = -U_p/(R_p C_p) + I/C_p \quad (3)$$

$$U_t = U_{OCV} - U_p - U_0 \quad (4)$$

According to Eq.2-4, the discretized state space equation is:

$$\begin{bmatrix} SOC_{k+1} \\ U_{p,k+1} \end{bmatrix} = \begin{bmatrix} 1 & 0 \\ 0 & \exp(-\Delta t/(R_p C_p)) \end{bmatrix} \cdot \begin{bmatrix} SOC_k \\ U_{p,k} \end{bmatrix} + \begin{bmatrix} -(\eta * \Delta t)/Q_n \\ R_p(1-\exp(-\Delta t/(R_p C_p))) \end{bmatrix} \cdot I_k \quad (5)$$

$$U_{t,k} = U_{OCV,k} - U_{p,k} - U_{0,k} \quad (6)$$

Where, the Eq.5 is the input equation, Eq.6 is the output equation, the state variable is $x_k = [U_{p,k} \ SOC_k]^T$, and the output is $U_{t,k}$. The Laplace transform of the Eq. 3-4 can be obtained:

$$U_{OCV}(s) - U_t(s) = (R_0 + R_p/(1+\tau_p s))*I(s) \quad (7)$$

where $\tau_p$ is the time constant and $\tau_p = R_p C_p$. Let $G(s) = (U_{OCV}(s) - U_t(s))/I(s)$, then Eq.7 is written as the transfer function:

$$G(s) = (\tau_p R_0 s + R_0 + R_p)/(1+\tau_p s) \quad (8)$$

Discretization of Eq.8 by bilinear transformation yields:

$$U_e(k) = d_0 I(k) + d_1 I(k-1) + d_2 U_e(k-1) \quad (9)$$

where,

$U_e = U_{OCV} - U_t$, $d_0 = ((R_0+R_p)\Delta t + 2*R_0 \tau_p)/(2*\tau_p + \Delta t)$, $d_1 = ((R_0+R_p)\Delta t - 2*R_0 \tau_p)/(2*\tau_p + \Delta t)$, $d_2 = (2*\tau_p - \Delta t)/(2*\tau_p + \Delta t)$.

Then the parameter vector $\Theta = [d_0 \ d_1 \ d_2]$, the data vector $\Phi = [I(k) \ I(k-1) \ U_e(k-1)]^T$. The parameters $[d_0 \ d_1 \ d_2]$ can be obtained by FFRLS algorithm, and then the model parameters $[R_0 \ R_p \ C_p]$ can be obtained. Since FFRLS is already a mature application of the algorithm, there is no longer disclosed too much detail.

## 3 The Algorithms of SOC Estimation

### 3.1 The Extended Kalman Filter algorithm

For the filtering problem of nonlinear systems, the commonly used method is to transform it into an approximate linear filtering problem by using linearization techniques. The core idea of EKF [5] is that for general nonlinear systems, the nonlinear functions $f(*)$ and $h(*)$ are first expanded into Taylor series around the state quantity $\hat{X}_k$ and the second-order and above terms are omitted to obtain an approximate linear model. Then the linear KF algorithm is used to complete the filtering estimation of the target. Table 1 is the implementation process of estimating SOC of lithium-ion battery based on EKF algorithm.

Table 1: Implementation steps of SOC estimation based on EKF algorithm

| Algorithm implementation steps |
|---|
| Step1-Parameter Initialization, $R\_x$, $Q\_x$, $P$ are initial values of measurement noise covariance matrix, process noise covariance matrix and error covariance matrix. $X_0 = [SOC_{Init} \ U_{p,Init}]^T, R\_x, Q\_x, P$ |
| Step2-Calculate state matrix $A$ and input matrix $B$. $A = \begin{bmatrix} 1 & 0 \\ 0 & \exp(-\Delta t/(R_p C_p)) \end{bmatrix}$ $B = \begin{bmatrix} -(\eta*\Delta t)/Q_n \\ R_p(1-\exp(-\Delta t/(R_p C_p))) \end{bmatrix}$ |
| Step3-State prediction. $X_{k|k-1} = [SOC_k \ U_{p,k}] = A*X_{k-1} + B*I_k$ |
| Step4-Observational prediction, $U_{OCV,k}$ is obtained by substituting $SOC_k$ into $OCV_{SOC}$ function. $U_{t,pre,k} = U_{OCV,k} - U_{p,k} - R_0 * I_k$ |
| Step5-The first-order linearized observation equation, $dOCV$ is obtained by substituting $SOC_k$ into $\partial OCV_{SOC}/\partial SOC$. $H = [dOCV \ 1]$ |
| Step6-Calculate terminal voltage estimation error. $U_{t,error,k} = U_{t,measure,k} - U_{t,pre,k}$ |
| Step7-The covariance matrix prediction. $P_{k|k-1} = (A*P_{k-1|k-1}*A^T) + Q\_x$ |
| Step8-The Kalman gain calculation. $K_k = P_{k|k-1}*H^T*(\text{inv}(H*P_{k|k-1}*H^T + R\_x))$ |
| Step9-State update. $X_{k|k} = X_{k|k-1} + K_k * U_{t,error,k}$ |
| Step10-Covariance matrix update, $E$ is unit matrix. $P_{k|k} = (E-(K_k*H))*P_{k|k-1}$ |

### 3.2 The H Infinite Filter Algorithm

The goal of HIF is to minimize the following cost function:

$$J = \frac{\sum_{k=0}^{N-1} \|y_k - \hat{y}_k\|_{S_k}^2}{\|x_0 - \hat{x}_0\|_{p_0^{-1}}^2 + \sum_{k=0}^{N-1}(\|w_k\|_{q_k^{-1}}^2 + \|v_k\|_{r_k^{-1}}^2)} \quad (10)$$

where $x_0$ is the initial value of the state, $\hat{x}_0$ is the estimated value of the state, $y_k$ is the target state, $\hat{y}_k$ is its estimated value, and the weight matrices $p_0$, $q_k$, $r_k$ and $S_k$ are symmetric positive definite matrices, which should be chosen according to the specific estimated object. The norm expression is defined as $\|m\|_\rho^2 = m^T \rho m$.

The core idea of HIF is to minimize the estimation error when the noise reaches the upper bound. However, it is very difficult to directly minimize the cost function $J$. Therefore, a performance boundary can be selected, and the method is used to estimate $\hat{y}_k$, so that the cost function $J$ satisfies: $J < 1/\gamma$. The limiting factor $\gamma$ is the specified performance boundary. Consider setting the custom limiting factor $\gamma$ to a fixed value. This can guarantee the optimized boundary constraints in advance. When $\gamma = 0$, HIF is equivalent to a KF, so the KF is a special case of HIF actually when its performance boundary is infinite.

To implement the HIF algorithm, $S_x$, $L_x$ and $\gamma$ need to be introduced, where $L_x$ is generally a unit matrix and $\gamma$ can be the value you need. Compared with the EKF algorithm, steps 2-7 remain invariable, and after step 7, it is updated to:

Step8-Symmetric positive definite matrix calculation.
$$S_{x,k|k-1} = L_x * S_{x,k-1} * L_x^T \tag{11}$$
Step9-The Kalman gain calculation.
$$K_k = P_{k|k-1} * (\text{inv}(E - \gamma * S_{x,k|k-1} * P_{k|k-1} + H^T * (\text{inv}(R\_x)) * H * P_{k|k-1})) * H^T * \text{inv}(R\_x) \tag{12}$$
Step10-State update.
$$X_{k|k} = X_{k|k-1} + K_k * U_{t,error,k} \tag{13}$$
Step11-Covariance matrix Update.
$$P_{k|k} = P_{k|k-1} * (\text{inv}(E - \gamma * S_{x,k|k-1} * P_{k|k-1} + H^T * (\text{inv}(R\_x)) * H * P_{k|k-1}) \tag{14}$$

### 3.3 Improved Adaptive H Infinity Extended Kalman Filter

This paper considers the adaptive updating process of the noise covariance matrix $Q\_x$ and the measurement noise covariance matrix $R\_x$ based on the battery model output voltage residual sequence. The updating process is as follows [7]:
$$\begin{cases} Q\_x_k = K_k * M * K_k^T \\ R\_x_k = M - H * P_{k|k-1} * H^T \end{cases} \tag{15}$$
$M$ is defined as follows:
$$M = \frac{1}{L} \sum_{j=k-L+1}^{k} (U_{t,error,j} * U_{t,error,j}^T) \tag{16}$$

Here, $U_{t,error}$ is the voltage residual of the battery model at $k$-th, $M$ is the approximate value of the voltage residual at $k$-th, and $L$ is the window length of the error covariance matching, which can be referred to [7, 14].

In order to improve the fast tracking of $Q\_x$ and $R\_x$, reduce the influence of historical residual data on the estimation update of noise covariance matrix at $k$-th, and avoid the failure of the algorithm caused by the negative number in the calculation process of $R\_x$, an improved AHIEKF is proposed, where the value range of $b \subseteq (0.9, 1)$:
$$\begin{cases} d = (1-b)/(1-b^k) \\ Q\_x_k = K_k * (d*M) * K_k^T \\ R\_x_k = (1-d)*M + H*P_{k|k-1}*H^T \end{cases} \tag{17}$$

## 4 The Results of Experiment

The experimental data in this paper comes from the SAMSUNG ICR18650-22P NCM power battery test. The rated capacity of the battery is 2.2 Ah, and the charge and discharge cut-off voltages are 4.2 V and 2.75 V, respectively. The power battery test conditions include capacity test, HPPC test, DST test and FUDS test. The test environment temperature is 25 °C and the sample time is 1s. According to the capacity test, the actual capacity of the battery is 2.0383 Ah.

### 4.1 The Function Relationship of OCV-SOC

According to the analysis of Section 2.1, the OCV is identified by DST condition based on the Rint model and the forgetting factor is set to 0.996 of FFRLS, and the initial value of OCV is set to 4. Fig.2 shows the OCV identification results, where the red curve is obtained by fitting the HPPC condition data points, and the blue curve is obtained by DST condition identification. It can be seen from the identification result that the identified OCV and the fitted OCV have the same trend and size. Subsequently, based on the seven parameters in FFRLS identification Eq.1, the final seven parameters are:

[$k_6$ $k_5$ $k_4$ $k_3$ $k_2$ $k_1$ $k_0$]=[-0.5061, 11.1208, -27.5840, 25.9496, -10.4888, 2.3296, 3.3398]

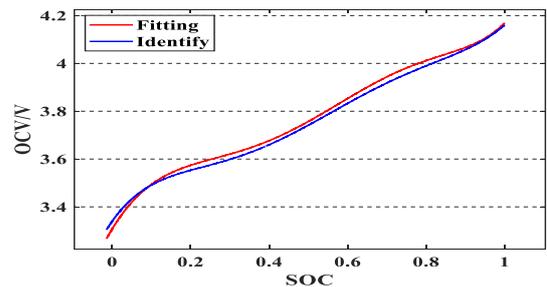

Fig.2: The result of OCV identification

## 4.2 Online Estimation of SOC

The simulation model is built in R2021b MATLAB/Simulink (Fig.3). The model is mainly composed of two parts: the part of model parameters identification and the part of lithium-ion battery SOC estimation. The final estimation results will demonstrate the superiority of this method from the Root-mean-square-error (*RMSE*) and Mean-absolute-error (*MAE*).

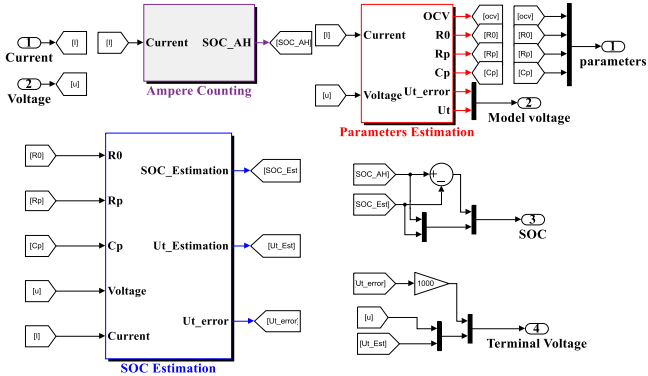

Fig.3: The simulation model of online SOC estimation

### 4.2.1 SOC Estimation Based on DST

The initial parameters are set as: $[SOC_0; Up_0]=[0.8; 0]$, the window length of $L$ is set as 5, $b=0.96$, $R\_x=0.8$, $Q\_x=[1e-5\ 0;\ 0\ 1e-5]$, $P=[0.035\ 0;\ 0\ 0.25]$, $S\_x=[0.9\ 0;\ 0\ 0.1]$, $\gamma=0.005$, the forgetting factor of model parameters identification is set as 0.999. According to the simulation model of Fig. 3, the model parameters identified at $k$-th are used for SOC estimation of lithium-ion battery in real time. Finally, the estimated SOC is compared with the reference SOC, and the *RMSE* and *MAE* error indexes are calculated.

Fig.4 is the SOC estimation results based on EKF, HIEKF, AHIEKF and IAHIEKF. Fig.5 is the SOC estimation error curve, and Table 2 is the error index of the four methods. The estimation accuracy orders of the four methods from high to low is IAHIEKF, AHIEKF, HIELF and EKF.

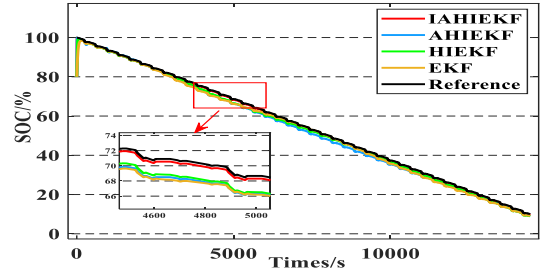

Fig.4: The results of SOC estimation based on DST

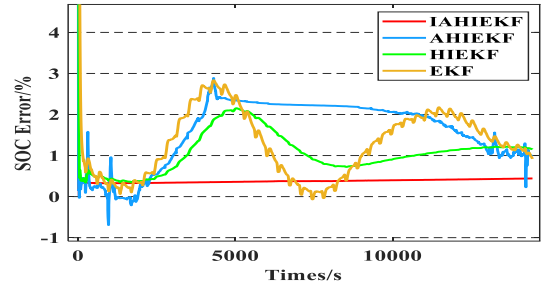

Fig.5: The errors of SOC estimation based on DST

Table 2: The estimation error indexes of DST condition

| Methods | *RMSE*/% | *MAE*/% |
|---------|----------|---------|
| IAHIEKF | 0.6008   | 0.3578  |
| AHIEKF  | 1.0896   | 0.8230  |
| HIEKF   | 1.6443   | 1.3100  |
| EKF     | 1.6444   | 1.3100  |

### 4.2.2 SOC Estimation Based on FUDS

The initial parameter setting is consistent with the DST condition. Compared with the DST condition, the FUDS condition has more severe current change and faster SOC reduction. Fig.6 is the SOC estimation results based on EKF, HIEKF, AHIEKF and IAHIEKF. Fig.7 is the SOC estimation error curves. Table 3 is the error index of the four methods. The estimation accuracy orders of the four methods from high to low is IAHIEKF, AHIEKF, HIELF and EKF.

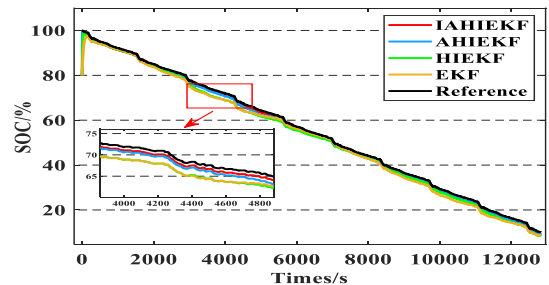

Fig.6: The results of SOC estimation based on FUDS

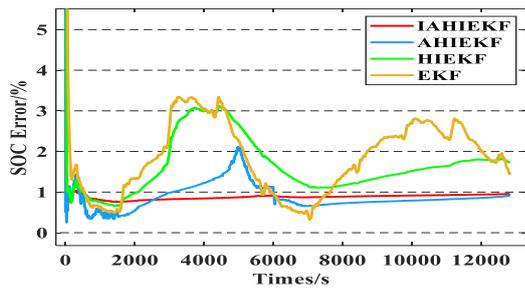

Fig.7: The errors of SOC estimation based on FUDS

Table 3: The estimation error indexes of FUDS condition

| Methods | RMSE/% | MAE/% |
|---|---|---|
| IAHIEKF | 1.0068 | 0.8721 |
| AHIEKF | 1.9778 | 1.8007 |
| HIEKF | 2.1643 | 1.8756 |
| EKF | 2.1643 | 1.8756 |

## 5 Conclusion

According to the error indicators in Table 2 and Table 3, it can be found that the IAHIEKF algorithm has the highest SOC estimation accuracy than the other three algorithms. The *RMSE* and *MAE* are 0.6008 % and 0.3578 % under DST condition, and 1.0068% and 0.8721 % under FUDS condition respectively. The results show that the proposed method has high accuracy and robustness for SOC estimation of lithium-ion battery, and has the ability of fast convergence when the initial value of SOC is not equal to the real value. Compared with the other three algorithms, IAHIEKF can maintain good estimation accuracy even if the working condition is changed. Stable and accurate SOC estimation has a great significance for estimating other battery states (Such as SOP and SOE).

## References

[1] IEA (2022), Global EV Outlook 2022, IEA, Paris https://www.iea.org/reports/global-ev-outlook-2022, License: CC BY 4.0

[2] Song X, Yang F, Wang D, et al. Combined CNN-LSTM network for state-of-charge estimation of lithium-ion batteries[J]. Ieee Access, 2019, 7: 88894-88902.

[3] Ng K S, Moo C S, Chen Y P, et al. Enhanced coulomb counting method for estimating state-of-charge and state-of-health of lithium-ion batteries[J]. Applied energy, 2009, 86(9): 1506-1511.

[4] Hossain M, Haque M E, Arif M T. Kalman filtering techniques for the online model parameters and state of charge estimation of the Li-ion batteries: A comparative analysis[J]. Journal of Energy Storage, 2022, 51: 104174.

[5] Xile D, Caiping Z, Jiuchun J. Evaluation of SOC estimation method based on EKF/AEKF under noise interference[J]. Energy Procedia, 2018, 152: 520-525.

[6] Zhao L, Liu Z, Ji G. Lithium-ion battery state of charge estimation with model parameters adaptation using H∞ extended Kalman filter[J]. Control Engineering Practice, 2018, 81: 114-128.

[7] Sun D, Yu X, Wang C, et al. State of charge estimation for lithium-ion battery based on an Intelligent Adaptive Extended Kalman Filter with improved noise estimator[J]. Energy, 2021, 214: 119025.

[8] Zhou W, Zheng Y, Pan Z, et al. Review on the battery model and SOC estimation method[J]. Processes, 2021, 9(9): 1685.

[9] Xiong R, Cao J, Yu Q, et al. Critical review on the battery state of charge estimation methods for electric vehicles[J]. Ieee Access, 2017, 6: 1832-1843.

[10] Xing Y, He W, Pecht M, et al. State of charge estimation of lithium-ion batteries using the open-circuit voltage at various ambient temperatures[J]. Applied Energy, 2014, 113: 106-115.

[11] Bester J E, El Hajjaji A, Mabwe A M. Modelling of Lithium-ion battery and SOC estimation using simple and extended discrete Kalman Filters for Aircraft energy management[C]//IECON 2015-41st Annual Conference of the IEEE Industrial Electronics Society. IEEE, 2015: 002433-002438.

[12] Chen X, Lei H, Xiong R, et al. A novel approach to reconstruct open circuit voltage for state of charge estimation of lithium-ion batteries in electric vehicles[J]. Applied Energy, 2019, 255: 113758.

[13] Susanna S, Dewangga B R, Wahyungoro O, et al. Comparison of simple battery model and thevenin battery model for SOC estimation based on OCV method[C]//2019 International Conference on Information and Communications Technology (ICOIACT). IEEE, 2019: 738-743.

[14] Maofei T, Zhiguo A N, Xing C, et al. SOC estimation of lithium battery based online parameter identification and AEKF[J]. Energy Storage Science and Technology, 2019, 8(4): 745.